\documentclass[reprint,aps,prl,superscriptaddress,lengthcheck]{revtex4-2}
\usepackage{epsfig}
\usepackage{latexsym}
\usepackage{xspace}
\usepackage[colorlinks=true,linktocpage=true,linkcolor=blue,citecolor=blue,allcolors=blue]{hyperref}
\usepackage[utf8]{inputenc}
\usepackage{indentfirst}
\usepackage{enumerate}
\usepackage{color}

\usepackage[caption=false,position=top]{subfig}

\usepackage{amsmath}
\usepackage{amssymb}
\usepackage[english]{babel}
\usepackage{url}

\newcommand{\bs}{\boldsymbol}

\newcommand{\eq}[1]{\begin{align} #1 \end{align}}
\newcommand{\mean}[1]{\langle #1 \rangle}
\newcommand{\cum}[1]{\kappa_{#1}}

\newcommand{\cumc}[1]{\hat{C}_{#1}}

\newcommand{\sNN}{\sqrt{s_{\rm NN}}}

\newcommand{\cc}[1]{\chi_{#1}^B}
\newcommand{\ccb}[1]{\bar{\chi}_{#1}^{B}}
\newcommand{\hd}[1]{\widehat{\Delta}\chi_{#1}^{B}}

\begin{document}

\title{
Extraction of baryon number susceptibilities at finite density from heavy-ion collisions
}

\author{Gr\'egoire Pihan}
    \affiliation{Physics Department, University of Houston, Houston, TX 77204, USA}

\author{Roman Poberezhniuk}
    \affiliation{Physics Department, University of Houston, Houston, TX 77204, USA}
    \affiliation{Bogolyubov Institute for Theoretical Physics, 03680 Kyiv, Ukraine}

\author{Volodymyr~A. Kuznietsov}
    \affiliation{Physics Department, University of Houston, Houston, TX 77204, USA}    
    \affiliation{Bogolyubov Institute for Theoretical Physics, 03680 Kyiv, Ukraine}

\author{Volodymyr~Vovchenko}
    \affiliation{Physics Department, University of Houston, Houston, TX 77204, USA}

\begin{abstract}
We present, to our knowledge, the first Bayesian extraction of baryon number susceptibilities of QCD matter at finite baryon density from heavy-ion collision data on proton number cumulants.
The framework embeds arbitrary equation-of-state susceptibilities $\chi_n^B$ into realistic hydrodynamic particlization hypersurfaces through maximum-entropy freeze-out, maps the resulting (anti)baryon fluctuations onto protons, applies the experimental kinematic acceptance, and accounts for exact baryon number conservation.
Applying the framework to measurements of (net-)proton number fluctuations in 0--5\% central Au-Au collisions from the RHIC Beam Energy Scan in the collider mode, we extract, at each collision energy, the second-order susceptibility normalized by the hadron resonance gas value, $\cc{2}/\ccb{2}$, and the higher-order susceptibility ratios $\cc{3}/\cc{1}$ and $\cc{4}/\cc{2}$.
We obtain tight constraints on $\cc{2}$, with extracted values in quantitative agreement with lattice QCD based estimates along the chemical freeze-out line for $\mu_B \lesssim 300$~MeV. 
At larger $\mu_B$, the extracted values indicate an enhancement of baryon number fluctuations relative to the noncritical lattice-based extrapolation and HRG baseline considered here.
The third- and fourth-order susceptibilities are only weakly constrained. 
In particular, we find that the observed nonmonotonic collision-energy dependence of the proton factorial cumulant ratio $\cumc{3}/\cumc{1}$
can be described without irreducible three- or four-baryon correlations.
This behavior emerges from the interplay between the energy dependence of $\cc{2}$ and exact baryon number conservation.
\end{abstract}

\maketitle

\emph{Introduction.---}
Fluctuations of conserved charges are sensitive probes of the QCD phase structure~\cite{Stephanov:2008qz,Bzdak:2019pkr}.
In equilibrium, they are quantified by baryon number susceptibilities 
$\cc{n} = \partial^n (P/T^4) / \partial (\mu_B/T)^n$,
which encode the response of strongly interacting matter to changes in the baryon chemical potential $\mu_B$. In particular, they are expected to develop characteristic non-monotonic structures near the conjectured QCD critical point~(CP)~\cite{Stephanov:2008qz,Stephanov:2011pb}.
While first-principles lattice QCD calculations determine $\cc{n}$ at $\mu_B = 0$~MeV~\cite{Bazavov:2017dus,Borsanyi:2018grb,Bollweg:2021vqf,Borsanyi:2023wno}, their behavior at finite baryon density relies on extrapolations~\cite{Bollweg:2022fqq,Basar:2023nkp,Clarke:2024ugt,Shah:2024img,Abuali:2025tbd} or alternative approaches~\cite{Lu:2025cls,Fischer:2026uni}, and remains a subject of active research. 
Experimentally, the QCD phase structure can be probed through event-by-event fluctuations of the net-proton number measured in relativistic heavy-ion collisions~\cite{Stephanov:1998dy,Stephanov:1999zu,Asakawa:2009aj,Athanasiou:2010kw}, most notably in the RHIC Beam Energy Scan~(BES) program, which covers baryochemical potentials up to \mbox{$\mu_B \sim 400$~MeV} in collider mode~\cite{Bzdak:2019pkr,STAR:2020tga}.
The high-precision BES-II measurements of (net-)proton cumulants and factorial cumulants in Au-Au collisions at $\sNN = 7.7$--$27$~GeV~\cite{STAR:2025zdq} exhibit intriguing deviations from non-critical baselines at the lowest energies.
The quantitative interpretation of these data, however, is obscured by effects not related to the equilibrium susceptibilities: the measured protons carry only part of the baryon number in the system, the acceptance is limited in momentum,
and exact baryon-number conservation non-trivially modifies the fluctuations relative to the grand-canonical expectation.
The usual strategy has been to confront the data with model calculations, such as the hadron resonance gas~(HRG) with exact baryon conservation~\cite{Braun-Munzinger:2020jbk}, hydrodynamics with excluded-volume~\cite{Vovchenko:2021kxx}, functional-renormalization-group~\cite{Zhao:2026mcp} susceptibilities, or transport models~\cite{Bass:1998ca,Bleicher:1999xi}.

In this Letter, we adopt a different strategy and perform, to our knowledge, the first Bayesian extraction of the baryon number susceptibilities from the data.
Rather than assuming a particular equation of state~(EoS), we infer the susceptibilities at freeze-out from the measured \mbox{(net-)proton} fluctuations, with the hydrodynamic background, baryon-to-proton mapping, experimental acceptance, and exact conservation incorporated explicitly.
The result is a data-driven determination of $\cc{2}$, $\cc{3}$, and $\cc{4}$ of hot QCD matter along the chemical freeze-out line, which can be directly confronted with lattice QCD and effective theories.

\emph{Framework.---} We construct a forward map from baryon susceptibilities on a hydrodynamic particlization hypersurface to the measured proton and net-proton cumulants. We then perform a Bayesian extraction of these susceptibilities from STAR data on proton fluctuations. 
The forward map proceeds in four steps:
(i) event-averaged hydrodynamic hypersurfaces are generated with \texttt{MUSIC}~\cite{Schenke:2010nt,Paquet:2015lta,Denicol:2018wdp}, using collision-geometry-based 3D initial conditions~\cite{Shen:2020jwv} and particlization at a constant energy density, \mbox{$\varepsilon_{\rm sw}=0.26$~GeV/fm$^3$}~\footnote{
Here we supplement Ref.~\cite{Shen:2020jwv} by performing additional MUSIC simulations at BES-II energies, $\sNN = 9.2$, $11.5$, and $17.3$~GeV.}, (ii) maximum-entropy freeze-out (MaxEnt)~\cite{Pradeep:2022eil,Karthein:2025hvl} converts the local baryon susceptibilities in each element of the hydrodynamic hypersurface $\Sigma$ into joint baryon–antibaryon cumulants, (iii) Cooper–Frye acceptance~\cite{Vovchenko:2021kxx} and baryon-to-proton filtering~\cite{Kitazawa:2011wh} map these cumulants onto those of accepted protons and antiprotons, and (iv) the subensemble acceptance method (SAM-3.0)~\cite{Poberezhniuk:2026bfv} imposes exact global baryon number conservation. 

At each collision energy, we parametrize the local baryon susceptibility input as
\eq{
\label{eq:gammas}
{\bs \gamma} = (\gamma_1,\gamma_2,\gamma_3)
=\left(
\frac{\cc{2}}{\ccb{2}},
\frac{\cc{3}}{\cc{1}},
\frac{\cc{4}}{\cc{2}}
\right).
}
Here $\ccb{2}$ is the second-order baryon susceptibility of the ideal HRG (iHRG), which serves as a normalization of $\cc{2}$. It represents the uncorrelated variance in the absence of genuine two-baryon correlations. In each hypersurface element $x\in\Sigma$, the iHRG equation of state~\cite{Vovchenko:2019pjl} is matched to the local hydrodynamic energy and net-baryon densities~\footnote{Therefore, by construction, $\cc{1} = \ccb{1}$.}, defining the Skellam baseline
\mbox{$\ccb{n}(x)=\bar{\chi}_{1}^{+}(x)+(-1)^n\bar{\chi}_{1}^{-}(x)$},
where \mbox{$\bar{\chi}_{1}^{+(-)}(x)=\rho_{B(\bar B)}(x)/T^3(x)$}. Here, $\rho_{B(\bar B)}(x)$ and $T(x)$ are the local iHRG baryon (antibaryon) densities and temperature obtained from this matching.
For ${\bs\gamma}=(1,1,1)$, all susceptibilities reduce to their local iHRG values. While $\cc{1}(x)$ and $\ccb{2}(x)$ vary across the hypersurface, the ratios ${\bs\gamma}$ are assumed uniform at each collision energy to facilitate their Bayesian extraction and should therefore be interpreted as effective hypersurface-averaged quantities.

The MaxEnt freeze-out prescription~\cite{Pradeep:2022eil} maps hydrodynamic fields, together with their fluctuations and correlations, onto hadron multiplicity irreducible relative cumulants~(IRCs) at freeze-out. The formalism accommodates arbitrary numbers of fields and particle species. 
In this Letter, we retain only local baryon density fluctuations, treat baryons and antibaryons as the two particle species, and regard each hypersurface element $x \in \Sigma$ as an independent grand-canonical subsystem. Under these assumptions, the \mbox{MaxEnt} particle IRCs coincide with the joint baryon–antibaryon factorial cumulants~\cite{Pradeep:2022eil}.
The first-order factorial cumulants are fixed by the local iHRG mean baryon and antibaryon densities. 
At higher orders, the MaxEnt method defines $\hd{k}(x)$, the order-$k$ irreducible susceptibility deviation relative to the local iHRG baseline. It excludes contributions reducible to lower-order deviations and yields the $(B^+, B^-)$ joint factorial cumulants, for $n+m\geq 2$, in the compact form
\eq{
\label{eq:maxent}
d \hat{C}_{nm}^{+-, \rm gce}(x) & =
d V_{\rm eff}(x)
[T(x)]^3 [w_+(x)]^n \nonumber \\
& \quad \times (-1)^m [w_-(x)]^m
\hd{n+m}(x),
}
where $d V_{\rm eff}(x) = d \Sigma_\mu(x) u^{\mu} (x)$, and $d \Sigma_\mu(x)$ and $u^{\mu} (x)$ are the oriented hypersurface element and the hydrodynamic four-velocity, respectively. The species weights, $w_+$ and $w_-$, are defined as 
\eq{
w_{\pm}(x) = \frac{\bar \chi_1^\pm(x)}{\bar \chi_1^+(x) + \bar \chi_1^-(x)}.
}
The MaxEnt prescription gives a direct connection between the deviations $\hd{n+m}$ in Eq.~\eqref{eq:maxent} and the regular susceptibility ratios Eq.~\eqref{eq:gammas}. For instance, at second order, 
\eq{
\hd{2} = \chi_2^B - \bar{\chi}_2^B = (\gamma_1 - 1) \ccb{2}.
}
The corresponding relations at third and fourth order are given in the Supplemental Material~\cite{SM}.

After MaxEnt determines the local joint baryon--antibaryon factorial cumulants, we map them onto the corresponding accepted proton and antiproton cumulants through independent binomial sampling. For each hypersurface element $x$ the sampling probability is 
\mbox{$\alpha_{p(\bar p)}(x)=q_\pm(x)p(x)$}. Here, 
$q_\pm(x)$ is the local proton-to-baryon or antiproton-to-antibaryon fraction in the final state calculated within the iHRG, and $p(x)$ is the Cooper--Frye probability for the particle to fall within the STAR acceptance.
Applying this binomial
thinning~\cite{Vovchenko:2021kxx,Kitazawa:2011wh} and summing the local contributions over the hypersurface yields the grand-canonical joint cumulants
$\kappa_{nm}^{\rm gce}(N_{\rm acc},B_{4\pi})$ with $n+m\leq4$. 
Here $N_{\mathrm{acc}}$ is the accepted
(net-)proton number and $B_{4\pi}$ is the total net baryon number. 
These joint cumulants are then provided as input to SAM-3.0~\cite{Poberezhniuk:2026bfv}, which imposes fixed $B_{4\pi}$ and yields the canonical cumulants of $N_{\rm acc}$ used in the comparison with data.
The explicit construction is given in the
Supplemental Material~\cite{SM}.

\emph{Bayesian inference.---}
At each collision energy we infer the posterior distribution of $(\cc{2}/\ccb{2},\cc{3}/\cc{1},\cc{4}/\cc{2})$
from the STAR (net-)proton fluctuation measurements in 0--5\% central Au-Au collisions~\cite{STAR:2020tga,STAR:2021iop,STAR:2025zdq}.
We use flat priors
$\cc{2}/\ccb{2} \in [0.5, 1.5]$,
$\cc{3}/\cc{1} \in [-15, 15]$~\footnote{We used a widened prior range of $[-60,60]$ for $\sNN = 200$~GeV due to large extracted uncertainty.}, and
$\cc{4}/\cc{2} \in [-300, 300]$,
and a Gaussian likelihood over the three measured ratios, with statistical and systematic uncertainties added in quadrature~\footnote{
Because the corresponding experimental covariance matrices are unavailable, correlations among the three measured ratios are neglected in the likelihood
Separately, we cross-validate the framework by using the posterior inferred from each observable set to predict the other set, see Fig.~\ref{fig:cross} in the Supplemental Material~\cite{SM}.}.

Two complementary extractions are performed:
(i)~from the proton factorial cumulant ratios $\cumc{2}/\cumc{1}$, $\cumc{3}/\cumc{1}$, $\cumc{4}/\cumc{1}$, and
(ii)~from the net-proton cumulant ratios $\cum{2}/\mean{p + \bar p}$, $\cum{3}/\cum{1}$, $\cum{4}/\cum{2}$~\footnote{We use the notation from Ref.~\cite{Bzdak:2019pkr} to denote ordinary and factorial cumulants while the STAR publication~\cite{STAR:2025zdq} uses the opposite notation.}.
The two data sets emphasize different aspects of the measured fluctuations: the net-proton cumulants retain the antiproton correlations that the proton factorial cumulants discard, while the factorial cumulants isolate the genuine multi-proton correlations.
Their mutual consistency provides a cross-check of the framework,
in particular given that the analysis of the \mbox{BES-I} data indicated that the applicability of hydrodynamics may be questionable for antiproton cumulants~\cite{Bzdak:2025rhp}.
The posterior is evaluated on a parameter grid,
and we quote marginal distributions and credible intervals throughout, with an example corner plot shown in the Supplemental Material~\cite{SM}.
In the present pilot study, the quoted uncertainties propagate only the experimental errors. 
The hydrodynamic input~(initial state, transport coefficients, $\varepsilon_{\rm sw}$, event-averaged surfaces) and the freeze-out prescription are held fixed, with the baseline setup described in Refs.~\cite{Shen:2020jwv,Vovchenko:2021kxx}.
Therefore, the quoted uncertainties are conditional on the fixed hydrodynamic and freeze-out setup and provide a lower-bound
estimate for the uncertainties expected in a global analysis.
We checked that the susceptibility ratios extracted from net-proton cumulant fits remain robust under the variation of the particlization switching energy density over $\varepsilon_{\rm sw} = 0.18$--$0.50$~GeV/fm$^3$.

\begin{figure*}[t]
    \centering
    \includegraphics[width=0.99\textwidth]{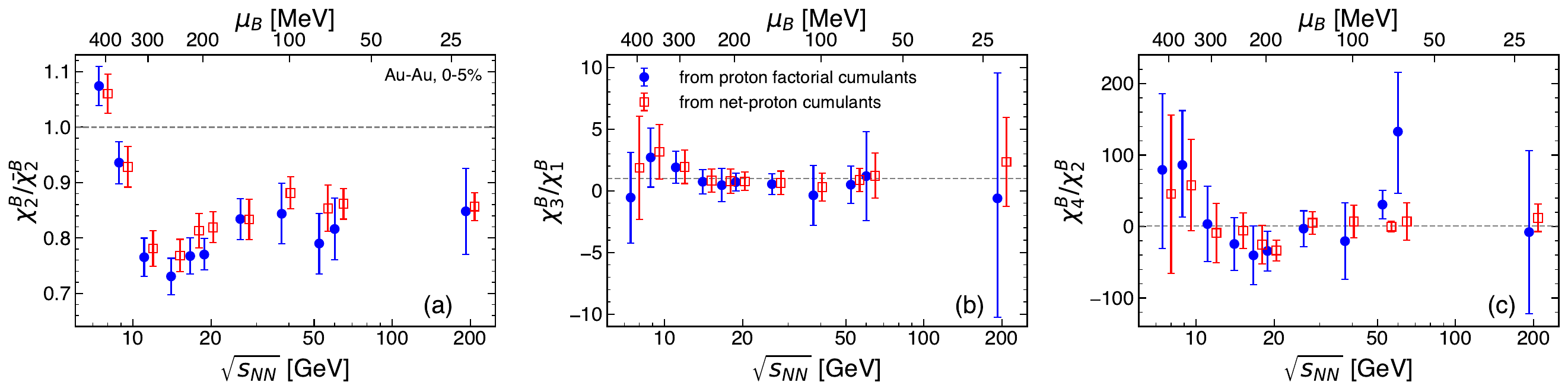}
    \caption{
    Collision energy dependence of the extracted baryon number susceptibility ratios $\cc{2}/\ccb{2}$~(a), $\cc{3}/\cc{1}$~(b), and $\cc{4}/\cc{2}$~(c) in 0--5\% central Au-Au collisions.
    Blue circles and red squares correspond to complementary extractions from proton factorial cumulants and net-proton cumulants~\cite{STAR:2025zdq,STAR:2020tga,STAR:2021iop}, respectively, slightly offset horizontally for clarity.
    Error bars indicate 68\% credible intervals of the marginal posterior distributions; bars spanning a large fraction of the prior range correspond to prior-dominated~(unconstrained) parameters.
    The upper axis indicates the baryochemical potential along the chemical freeze-out line~\cite{Vovchenko:2015idt}.
    }
    \label{fig:extraction}
\end{figure*}

\emph{Extracted susceptibilities.---}
Figure~\ref{fig:extraction} shows the central result of this Letter: the collision energy dependence of the extracted susceptibility ratios from both data sets.
The two complementary extractions agree within uncertainties at every energy, indicating internal consistency of the measured cumulants and factorial cumulants within the present framework.
Using net-proton cumulant data yields smaller uncertainties at higher $\sNN$ than proton factorial cumulants, as the former retain the information carried by antiprotons.

The analysis yields tight constraints on the value of the second-order susceptibility relative to iHRG~($\cc{2}/\ccb{2}$), with relative uncertainties within $5$-$10\%$.
Its energy dependence is non-monotonic:
an enhancement of \mbox{$\cc{2}/\ccb{2} \simeq 1.06$--$1.07$} relative to iHRG at the lowest collider energy of $\sNN = 7.7$~GeV, a minimum \mbox{$\cc{2}/\ccb{2} \simeq 0.73$--$0.82$} at $\sNN = 11.5$--$19.6$~GeV, and an approximately flat suppression $\cc{2}/\ccb{2} \simeq 0.8$--$0.9$ at high energies $\sNN \geq 39$~GeV.
While our analysis here is agnostic as to the physical mechanism behind the extracted behavior of $\cc{2}/\ccb{2}$, we note that the suppression relative to the Skellam baseline at higher energies is consistent with repulsive baryon interactions as described by the Hydro EV baseline~\cite{Vovchenko:2021kxx}.
On the other hand, the enhancement of fluctuations at the lowest energy is potentially intriguing and may signal the onset of attractive correlations at large $\mu_B$~\cite{Vovchenko:2021kxx,Friman:2025swg}. Its possible connection to critical behavior remains to be established.

The third- and fourth-order susceptibilities carry considerably larger uncertainties. 
The intermediate steps between baryon susceptibilities and measured protons, such as the baryon-to-proton mapping and the kinematic cuts, act analogously to an efficiency loss~\cite{Kitazawa:2011wh,Bzdak:2012ab}, diluting the sensitivity to higher-order correlations. 
The baryon conservation correction further distorts the sensitivity to $\cc{3}$ and $\cc{4}$.
The large uncertainties preclude definitive conclusions about the QCD phase structure from the high-order fluctuations, as the extracted values can accommodate scenarios ranging from a non-monotonic collision energy dependence of $\cc{4}/\cc{2}$ indicative of a QCD CP~\cite{Stephanov:2011pb} to the Skellam baseline of the ideal gas.
Nevertheless, meaningful constraints emerge at intermediate energies.
In particular, the fourth-order ratio exhibits its most significant deviation from zero at $\sNN = 19.6$~GeV: $\cc{4}/\cc{2} = -33 \pm 14$ from the net-proton fit,
indicating a local $2.3\sigma$ preference for a negative $\chi_4^B / \chi_2^B$,
in line with the deviations from the non-critical baseline at the same energy highlighted by STAR~\cite{STAR:2025zdq}.
The full set of extracted values is tabulated in the Supplemental Material~\cite{SM}.

\emph{Two-baryon correlation dominance.---}
\begin{figure}[t]
    \centering
    \includegraphics[width=0.49\textwidth]{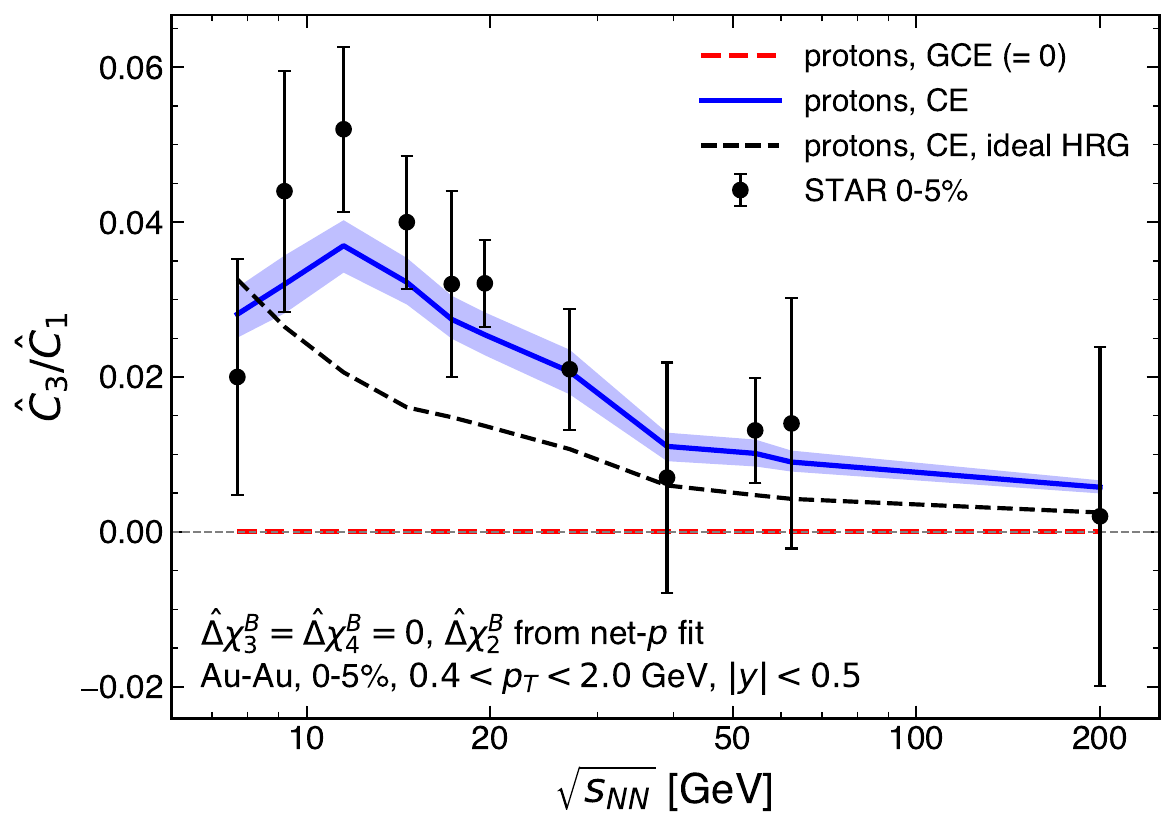}
    \caption{
    Collision energy dependence of the proton factorial cumulant ratio $\cumc{3}/\cumc{1}$ in the scenario retaining only the irreducible two-baryon correlation, $\hd{3} = \hd{4} = 0$, with $\cc{2}/\ccb{2}$ taken from the net-proton fit.
    The solid blue line with the 68\% band shows the full result for accepted protons with exact baryon conservation~(CE) and the dashed red line shows the vanishing ratio in the absence of the conservation correction~(GCE).
    The black dashed line shows the Skellam baseline ${\bs \gamma} = (1,1,1)$ with exact conservation.
    STAR data~\cite{STAR:2025zdq,STAR:2020tga,STAR:2021iop} are shown by symbols. 
    }
    \label{fig:stages}
\end{figure}
Given the weak constraints on $\cc{3}$ and $\cc{4}$, it is
natural to ask whether the data require any irreducible
multi-baryon correlations beyond second order. We therefore
consider a minimal scenario in which the EoS contains only an
irreducible two-baryon contribution, with the irreducible
three- and four-baryon contributions set to zero,
$\hd{3}=\hd{4}=0$.
The single parameter $\cc{2}/\ccb{2}$ is taken from the
net-proton fit at each energy.
In this scenario, the higher-order susceptibilities are not at
their Skellam values but are induced by the two-particle
correlations,
$
\Delta\chi_3^B=3\Delta f\,\Delta\chi_2^B$, 
$
\Delta\chi_4^B=
\left[4+3(\Delta f)^2\right]\Delta\chi_2^B,
$
where
$\Delta f(x)\equiv w_+(x)-w_-(x)
=\bar\chi_1^B(x)/\bar\chi_2^B(x)$, see Supplemental Material~\cite{SM}. These relations are
evaluated locally in each hypersurface element.

Figure~\ref{fig:stages} shows the collision-energy dependence of $\cumc{3}/\cumc{1}$ compared with the STAR data. The full set of factorial cumulant ratios is provided in the Supplemental Material~\cite{SM}. 
Our calculation reproduces the characteristic nonmonotonic peak observed in the data near $\sNN \approx 11$~GeV. 
This structure results from the interplay between exact baryon-number conservation and two-particle correlations: the nontrivial energy dependence of $\cc{2}/\ccb{2}$ modifies the nonzero $\cumc{3}/\cumc{1}$ generated by baryon conservation~\cite{Bzdak:2025rhp}.

This result demonstrates that the salient non-monotonic structure of the measured $\hat C_3/\hat C_1$ can be generated without irreducible three- or four-baryon correlations.
We note that this does not rule out sizable irreducible multi-baryon correlations in the EoS.
Rather, it reflects that the measurements are primarily sensitive to the irreducible two-baryon contribution.
Verifying the presence of irreducible multi-baryon correlations expected from critical behavior will thus require either a substantial reduction of experimental uncertainties or observables with enhanced sensitivity to genuine multi-particle correlations.

\begin{figure}[t]
    \centering
    \includegraphics[width=0.99\columnwidth]{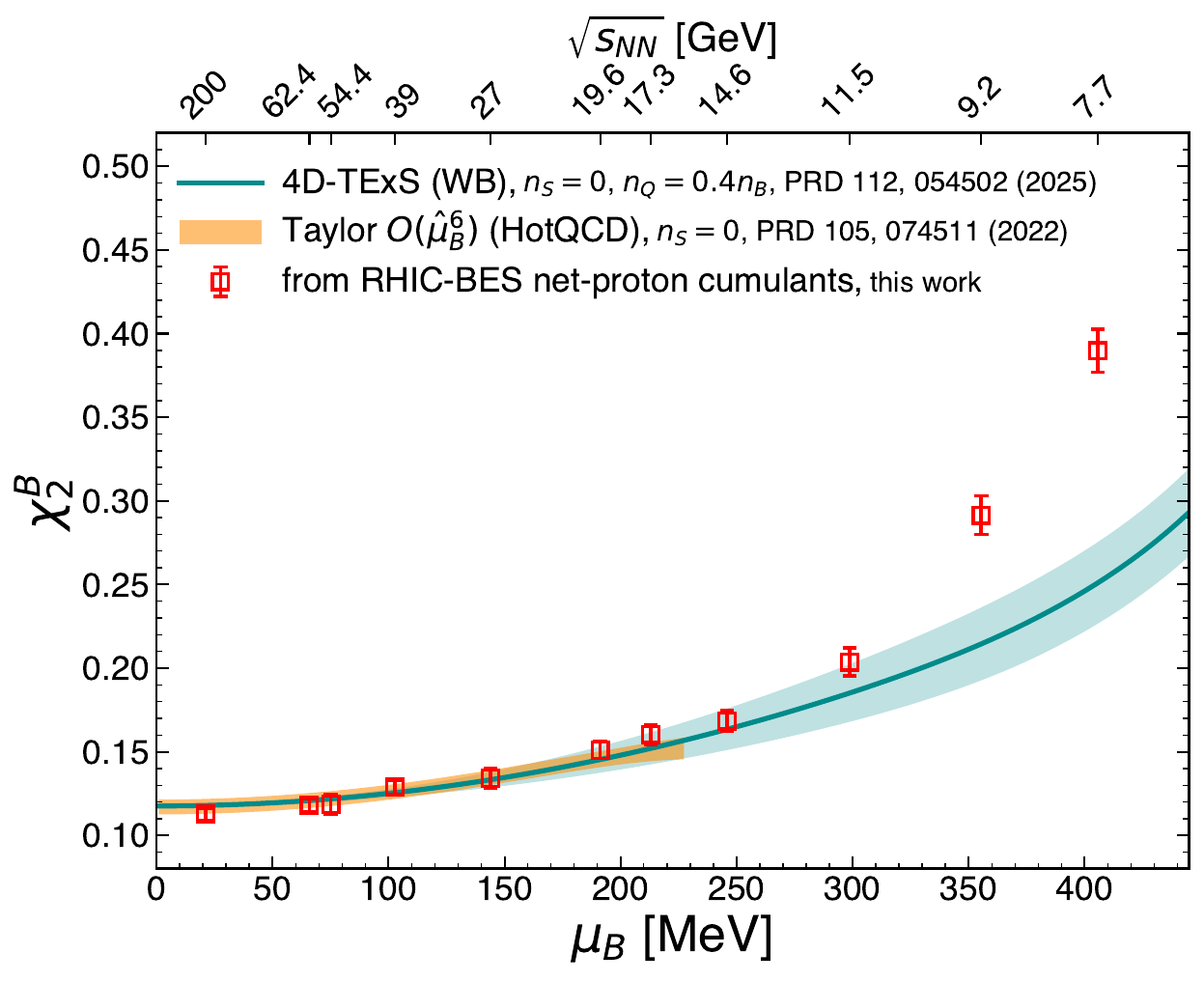}
    \caption{
    The second-order baryon number susceptibility $\cc{2}$ along the chemical freeze-out line~\cite{Vovchenko:2015idt} extracted from the net-proton cumulant data (red symbols), compared to the lattice-QCD-based estimates, including 4D-TExS~\cite{Abuali:2025tbd}~(teal curve with band) and the strangeness-neutral HotQCD Taylor expansion result~\cite{Bollweg:2022rps}~(orange band).
    The absolute normalization is obtained using the HRG model with the QMHRG2020 hadron list.
    }
    \label{fig:chi2B}
\end{figure}

\emph{Comparison with lattice QCD.---}
Next, we confront our data-driven determination of baryon number susceptibilities with first-principles lattice QCD estimates.
We use the chemical freeze-out line from Ref.~\cite{Vovchenko:2015idt} to map collision energies to $(T,\mu_B)$ points on the QCD phase diagram~\footnote{We obtain essentially the same result using an alternative freeze-out line from~\cite{Andronic:2017pug}.}.
We multiply the extracted $\cc{2}/\ccb{2}$ values by $\ccb{2}$ computed in the HRG model with the lattice-motivated QMHRG2020 hadron list~\cite{Bollweg:2021vqf} to obtain the absolute values of $\cc{2}$ and plot these as a function of $\mu_B$ in Fig.~\ref{fig:chi2B}.
We compare these values with a lattice-based EoS at finite density~\cite{Abuali:2025tbd}, evaluated within MUSES 4D-TExS code~\cite{Jahan:2026hvs,MUSES_4DTExS_v1_0_4}~(teal band), as well as the strangeness-neutral Taylor expansion result of the HotQCD collaboration~\cite{Bollweg:2022rps}~(orange band).
This comparison tests the hypothesis that the effective fluctuation freeze-out conditions track the conventional chemical freeze-out line.
We observe quantitative agreement, within uncertainties, between the data-driven and lattice-based values of $\cc{2}$ at $\mu_B \lesssim 300$~MeV.
The agreement is consistent with local equilibration of baryon number fluctuations.
At $\mu_B \gtrsim 300$~MeV, however, the extracted $\cc{2}$ deviates upward from the lattice band, with a relative enhancement reaching up to about $50\%$ at $\sNN = 7.7$~GeV.

We note the caveats entering this comparison: the absolute normalization of $\cc{2}$ depends on the hadron list, the mapping to the phase diagram assumes a specific freeze-out line for fluctuations, and the 4D-TExS EoS at these $\mu_B$ values is an extrapolation without a truncation error included.
Since the STAR data incorporate the centrality-bin-width
correction~\cite{Luo:2013bmi}, our calculation does not explicitly include the corresponding effects of volume fluctuations, nor those of initial-state and baryon-stopping fluctuations.
These effects are expected to become increasingly important at the lowest collision energies.
If more complete volume-fluctuation
corrections~\cite{Wang:2025fve} lead to appreciable revisions of the measured cumulants, the present extraction should be revisited.
We also note that an extended treatment of charge conservation effects, such as multiple conserved charges~\cite{Vovchenko:2020gne,Vovchenko:2022syc} or local baryon conservation~\cite{Braun-Munzinger:2023gsd,Vovchenko:2024pvk}, generally introduces further suppression of particle fluctuations.
Within the present framework, such suppression would tend to require larger values of the extracted $\cc{2}$ to reproduce the same data.
With these caveats in mind, the trend suggests a tension with the lattice-based extrapolation and warrants scrutiny with future fixed-target data.

\emph{Summary and outlook.---}
We performed a Bayesian extraction of baryon number susceptibilities of QCD matter at finite baryon density from heavy-ion collision data, based on relativistic hydrodynamics with maximum entropy freeze-out.
Within the current hydrodynamic and particlization setup, the experimental data constrain $\cc{2}/\ccb{2}$ at the few-percent level.
The extracted second-order susceptibility agrees quantitatively with the lattice-based 4D-TExS equation of state at $\mu_B \lesssim 300$~MeV, while exhibiting a sizable enhancement at larger $\mu_B$.
The third- and fourth-order susceptibilities remain weakly constrained within the current data precision. 
We find that the data can largely be described without
irreducible three- or four-baryon correlations, retaining only the irreducible two-baryon contribution.
In particular, the nonmonotonic energy dependence of the third-order proton factorial cumulant observed by STAR can be generated by the interplay between the energy dependence of $\cc{2}$ and exact baryon number conservation.

The framework presented here is general and opens several directions: the inclusion of fixed-target energies once hydrodynamic modeling and volume-fluctuation corrections are under control,
the treatment of energy density fluctuations and other conserved charges,
the incorporation of model uncertainties into the Bayesian analysis,
applications at the LHC and at other centralities and acceptances, and the use of the extracted susceptibilities as direct constraints on the finite-density QCD EoS.

\begin{acknowledgments}

\emph{Acknowledgments.} 
We thank Volker Koch, Chun Shen, and Misha Stephanov for fruitful discussions.
VV appreciates the hospitality within the nuclear theory group at LBNL, where part of this work was done.
This work was supported by the U.S. Department of Energy, Office of Science, Office of Nuclear Physics, Early Career Research Program under Award Number DE-SC0026065.
This work used computational resources of the Carya cluster
operated by the HPE DSI at the University of Houston.

\end{acknowledgments}

\bibliography{main}

\begin{appendix}
\setcounter{equation}{0}
\setcounter{figure}{0}
\renewcommand{\theequation}{A\arabic{equation}}
\renewcommand{\thefigure}{A\arabic{figure}}

\section{\large \bf Supplemental Material}
\section{MaxEnt mapping from baryon susceptibilities to
baryon--antibaryon cumulants}

For each hypersurface element $x\in\Sigma$, the iHRG state
matched to the local hydrodynamic energy and net-baryon densities
fixes the mean baryon and antibaryon susceptibilities
$\bar\chi_1^\pm(x)$. They define the local Skellam reference
\begin{equation}
\bar\chi_k^B(x)
=
\bar\chi_1^+(x)+(-1)^k\bar\chi_1^-(x)
\end{equation}
and the species weights
\begin{equation}
w_\pm(x)
=
\frac{\bar\chi_1^\pm(x)}
{\bar\chi_1^+(x)+\bar\chi_1^-(x)}.
\end{equation}

We retain only local fluctuations of the baryon-density mode and
treat baryons and antibaryons as the two particle species. Under these assumptions, the MaxEnt irreducible relative cumulants~(IRCs) coincide with the joint baryon--antibaryon factorial
cumulants. The first-order factorial susceptibilities are fixed
by the local iHRG means,
\begin{equation}
\label{eq:maxent0}
\hat\chi_{10}^{+-}(x)=\bar\chi_1^+(x),
\qquad
\hat\chi_{01}^{+-}(x)=\bar\chi_1^-(x).
\end{equation}

In the MaxEnt mapping~\cite{Pradeep:2022eil,Karthein:2025hvl}, retaining only the baryon-density mode
assigns to each particle index a projection factor given by its
ideal-gas covariance with that mode, normalized by the ideal
baryon-density variance. These factors are $w_+$ for a baryon
and $-w_-$ for an antibaryon. Therefore, for $n+m\geq2$,
\begin{equation}
\label{eq:maxentSM}
\hat\chi_{nm}^{+-}(x)
=
(-1)^m[w_+(x)]^n[w_-(x)]^m
\Delta\hat\chi_{n+m}^B(x).
\end{equation}
Thus, the contribution of the element to the corresponding joint
factorial cumulant is
\begin{equation}
d\hat C_{nm}^{+-,\mathrm{gce}}(x)
=
dV_{\mathrm{eff}}(x)[T(x)]^3
\hat\chi_{nm}^{+-}(x).
\end{equation}

Here $\Delta\hat\chi_k^B$ denotes the order-$k$ irreducible
susceptibility deviation of the baryon-density mode relative to
the iHRG reference. 
Reducing the general construction of Refs.~\cite{Pradeep:2022eil,Karthein:2025hvl}~(see Eq.~(32) in \cite{Karthein:2025hvl}) to the case of two species and single hydrodynamic field here, 
these quantities
are related through fourth order to the ordinary susceptibility
deviations by
\begin{align}
\Delta\hat\chi_2^B
&=
\Delta\chi_2^B,
\\
\Delta\hat\chi_3^B
&=
\Delta\chi_3^B
-
3\Delta f\,\Delta\chi_2^B,
\\
\Delta\hat\chi_4^B
&=
\Delta\chi_4^B
-
6\Delta f\,\Delta\chi_3^B
-
\left[4-15(\Delta f)^2\right]\Delta\chi_2^B.
\end{align}
Here $\Delta f = w_+ - w_-$.
In terms of the parametrization in Eq.~\eqref{eq:gammas} of the main text, one has
\begin{align}
\label{eq:devsSM}
\Delta\chi_2^B
&=
(\gamma_1-1)\bar\chi_2^B,
\\
\label{eq:devv3SM}
\Delta\chi_3^B
&=
(\gamma_2-1)\bar\chi_1^B,
\\
\label{eq:devv4SM}
\Delta\chi_4^B
&=
(\gamma_1\gamma_3-1)\bar\chi_2^B.
\end{align}
These relations guarantee that the net-baryon combinations of the resulting baryon--antibaryon cumulants in Eq.~\eqref{eq:maxentSM} reproduce the prescribed
susceptibilities locally in every hypersurface element.

\section{Joint cumulants of accepted (anti)protons and the conserved baryon number}
\label{sec:pipeline}

The SAM-3.0 conservation correction takes as input the grand-canonical joint cumulants $\kappa_{nm}^{\rm gce}(N_{\rm acc}, B_{4\pi})$ with $n + m \leq 4$, where $N_{\rm acc}$ is the accepted proton, antiproton, or net-proton number and $B_{4\pi} = N_+ - N_-$ is the total net baryon number.
The conserved $B_{4\pi}$ is fixed to the total net baryon number obtained by integrating the net baryon density over the particlization hypersurface.
In the grand-canonical calculation, different hypersurface elements are statistically independent, so the joint cumulants are additive,
\eq{\label{eq:jointtotgce}
\kappa_{nm}^{\rm gce}(N_{\rm acc}, B_{4\pi}) = \int_{\Sigma} d\kappa_{nm}(x),
}
and it suffices to specify the contribution $d\kappa_{nm}(x)$ of a single element.
Here we detail the construction of $d\kappa_{nm}(x)$ from the MaxEnt input~\eqref{eq:maxentSM} and the acceptance probabilities $\alpha_{p (\bar p)}(x) = q_{\pm}(x)\, p(x)$ defined in the main text, following the extended Cooper-Frye particlization procedure for fluctuations~\cite{Vovchenko:2020kwg,Vovchenko:2021kxx}.
The (anti)proton fractions $q_{\pm}(x)$ account for the feed-down from strong, electromagnetic, and weak decays, evaluated in the iHRG model with PDG2020 decay channels at the local thermodynamic conditions. 
The treatment of weak feed-down follows the baseline analysis of Ref.~\cite{Vovchenko:2021kxx}.

It is convenient to work with the ordinary joint cumulants of the (anti)baryon numbers emitted from the element,
\eq{
d\kappa^{+-}_{ij}(x) = dV_{\rm eff}(x)\, [T(x)]^3\, \chi^{+-}_{ij}(x),
}
where the ordinary joint susceptibilities $\chi^{+-}_{ij}$ follow from the factorial ones~[Eq.~\eqref{eq:maxentSM}] through the standard linear conversion relating ordinary and factorial cumulants,
\eq{\label{eq:stirlingSM}
\chi^{+-}_{ij} = \sum_{a \leq i} \sum_{b \leq j} S(i,a)\, S(j,b)\, \hat{\chi}^{+-}_{ab},
}
with $S(i,a)$ the Stirling numbers of the second kind and $S(i,0) \equiv \delta_{i0}$.
The conversion preserves the MaxEnt matching to the prescribed
net-baryon susceptibilities. Indeed, since $B=N_+-N_-$,
\eq{
\chi_k^B
=
\sum_{i=0}^{k}
\binom{k}{i}(-1)^{k-i}
\chi^{+-}_{i,k-i},
}
which is satisfied identically upon substituting
Eq.~\eqref{eq:stirlingSM} together with the MaxEnt factorial susceptibilities in Eqs.~\eqref{eq:maxent0} and \eqref{eq:maxentSM}.

For a given hypersurface element, let $dN_+$ and $dN_-$ denote the random baryon and antibaryon multiplicities emitted from that element. 
Conditional on $dN_{\pm}$, the accepted proton and antiproton multiplicities are independently binomially distributed,
\eq{
dn_p \sim \text{Bin}(dN_+,\alpha_p), \quad dn_{\bar p} \sim \text{Bin}(dN_-,\alpha_{\bar p}).
}
We define the local contributions to the accepted net-proton number and the conserved net-baryon number as
\eq{
d N_{\rm acc} = dn_p - dn_{\bar p}, \quad dB = dN_+ - dN_-.
}

The proton-only case used in the factorial-cumulant fit is obtained by setting $\alpha_{\bar p} = 0$.
The local joint cumulant generating function is
\eq{
K(t, t_B) \equiv \ln \mean{e^{t dN_{\rm acc} + t_B dB} }.
}
In terms of the cumulant-generating function of $(dN_+, dN_-)$,
\eq{
F(u_+, u_-) = \ln \mean{ e^{u_+ dN_+ + u_- dN_-} },
}
it is given by the exact composition
\eq{\label{eq:cgfSM}
K(t, t_B) = F\big(t_B + \phi_p(t),\, -t_B + \phi_{\bar p}(-t)\big),
}
where
\eq{
\phi_{a}(t) = \ln\left[ 1 - \alpha_{a} + \alpha_{a}\, e^{t} \right], \quad a = p, \bar{p},
}
is the cumulant generating function of a single Bernoulli trial, with cumulants $c_r(\alpha) \equiv \left. \tfrac{d^r}{dt^r} \ln\left(1-\alpha+\alpha e^{t}\right) \right|_{t=0}$, i.e. $c_1 = \alpha$, $c_2 = \alpha(1-\alpha)$, $c_3 = \alpha(1-\alpha)(1-2\alpha)$, etc.
Applying the multivariate Fa\`a di Bruno formula~\cite{Constantine:1996Faa,Vovchenko:2021yen,Poberezhniuk:2026bfv} to Eq.~\eqref{eq:cgfSM} gives the local joint cumulants, $d\kappa_{nm} = \left. \partial_t^n \partial_{t_B}^m K \right|_{t = t_B = 0}$, at arbitrary order:
\eq{\label{eq:masterSM}
d\kappa_{nm}
& =
\sum_{a=0}^{m}
\sum_{k=0}^{n}
\binom{m}{a}
\binom{n}{k}
(-1)^{m-a+n-k} \nonumber \\
& \quad \times
\sum_{r=0}^{k}
\sum_{s=0}^{n-k}
d\kappa^{+-}_{a+r,\,m-a+s}\,
B^{(p)}_{k,r}\,
B^{(\bar p)}_{n-k,s}.
}
Here $B^{(p)}_{k,r} \equiv B_{k,r}[c_1(\alpha_p), \ldots, c_{k-r+1}(\alpha_p)]$ and $B^{(\bar p)}_{l,s} \equiv B_{l,s}[c_1(\alpha_{\bar p}), \ldots, c_{l-s+1}(\alpha_{\bar p})]$ are the partial exponential Bell polynomials~\cite{Comtet:1974} evaluated on the Bernoulli cumulants, with $B_{0,0} = 1$ and $B_{k,0} = 0$ for $k \geq 1$.

For illustration, the lowest-order relations include
\begin{align}
d\kappa_{10}
&=
\alpha_p d\kappa_{10}^{+-}
-
\alpha_{\bar p}d\kappa_{01}^{+-},
\\
d\kappa_{01}
&=
d\kappa_{10}^{+-}
-
d\kappa_{01}^{+-},
\end{align}
while the cumulants of the conserved charge itself are independent
of the acceptance probabilities, for example,
\begin{equation}
d\kappa_{02}
=
d\kappa_{20}^{+-}
-
2d\kappa_{11}^{+-}
+
d\kappa_{02}^{+-}
=
dV_{\rm eff}T^3\chi_2^B.
\end{equation}
Higher-order expressions follow directly from Eq.~\eqref{eq:masterSM}.
In the proton-only limit $\alpha_{\bar p} = 0$, the diagonal cumulants $d\kappa_{n0}$ through fourth order reduce to the standard binomial-thinning~(efficiency-correction) formulas for proton cumulants~\cite{Kitazawa:2011wh,Bzdak:2012ab}; more generally, $d\kappa_{0m} = dV_{\rm eff}\, T^3\, \cc{m}$ at all orders, since the acceptance does not affect $B_{4\pi}$.
All expressions are exact for the given local input and are summed over the hypersurface to obtain the full grand-canonical result via Eq.~\eqref{eq:jointtotgce}.

Exact global conservation of $B_{4\pi}$ is imposed at the last step with the subensemble acceptance method \mbox{SAM-3.0~\cite{Poberezhniuk:2026bfv}}, which corrects the cumulants of $N_{\rm acc}$ order by order using the complete set of joint cumulants $\kappa_{nm}^{\rm gce}(N_{\rm acc}, B_{4\pi})$, $n + m \leq 4$.
In contrast to earlier SAM formulations~\cite{Vovchenko:2020tsr,Vovchenko:2021yen}, SAM-3.0 does not assume a uniform system or a particular relation between the accepted and total fluctuations and is therefore applicable to the present inhomogeneous hydrodynamic setting with arbitrary local susceptibilities.

The dependence of the accepted joint cumulants on the three susceptibility-deviation combinations $\gamma_1 - 1$, $\gamma_2 - 1$, and $\gamma_1 \gamma_3 - 1$ is linear:
\eq{\label{eq:templatesSM}
\kappa_{nm}^{\rm gce}({\bs \gamma}) & = \kappa_{nm}^{\rm ref}
+ (\gamma_1 - 1) \, \kappa_{nm}^{(2)} + (\gamma_2 - 1) \, \kappa_{nm}^{(3)} \nonumber \\
& \quad + (\gamma_1 \gamma_3 - 1) \, \kappa_{nm}^{(4)},
}
where $\kappa_{nm}^{\rm ref}$ is the Skellam reference and $\kappa_{nm}^{(k)}$ are precomputed response coefficients corresponding to a unit deviation at order $k$~[cf.~Eqs.~\eqref{eq:devsSM}--\eqref{eq:devv4SM}].
Only the~(nonlinear) SAM-3.0 correction is evaluated per parameter point.
The posterior is evaluated on a uniform grid spanning the prior ranges.
The likelihood is Gaussian in the three measured ratios with statistical and systematic uncertainties added in quadrature and correlations among the ratios neglected.
For the two-baryon-correlation scenario, $\hd{3} = \hd{4} = 0$, the induced susceptibility differences $\Delta \cc{3} = 3 \Delta f \, \Delta \cc{2}$ and $\Delta \cc{4} = [4 + 3 (\Delta f)^2] \, \Delta \cc{2}$ are implemented exactly at the cell level.

\section{Posterior distributions and cross-validation}

\begin{figure}[b]
    \centering
    \includegraphics[width=0.49\textwidth]{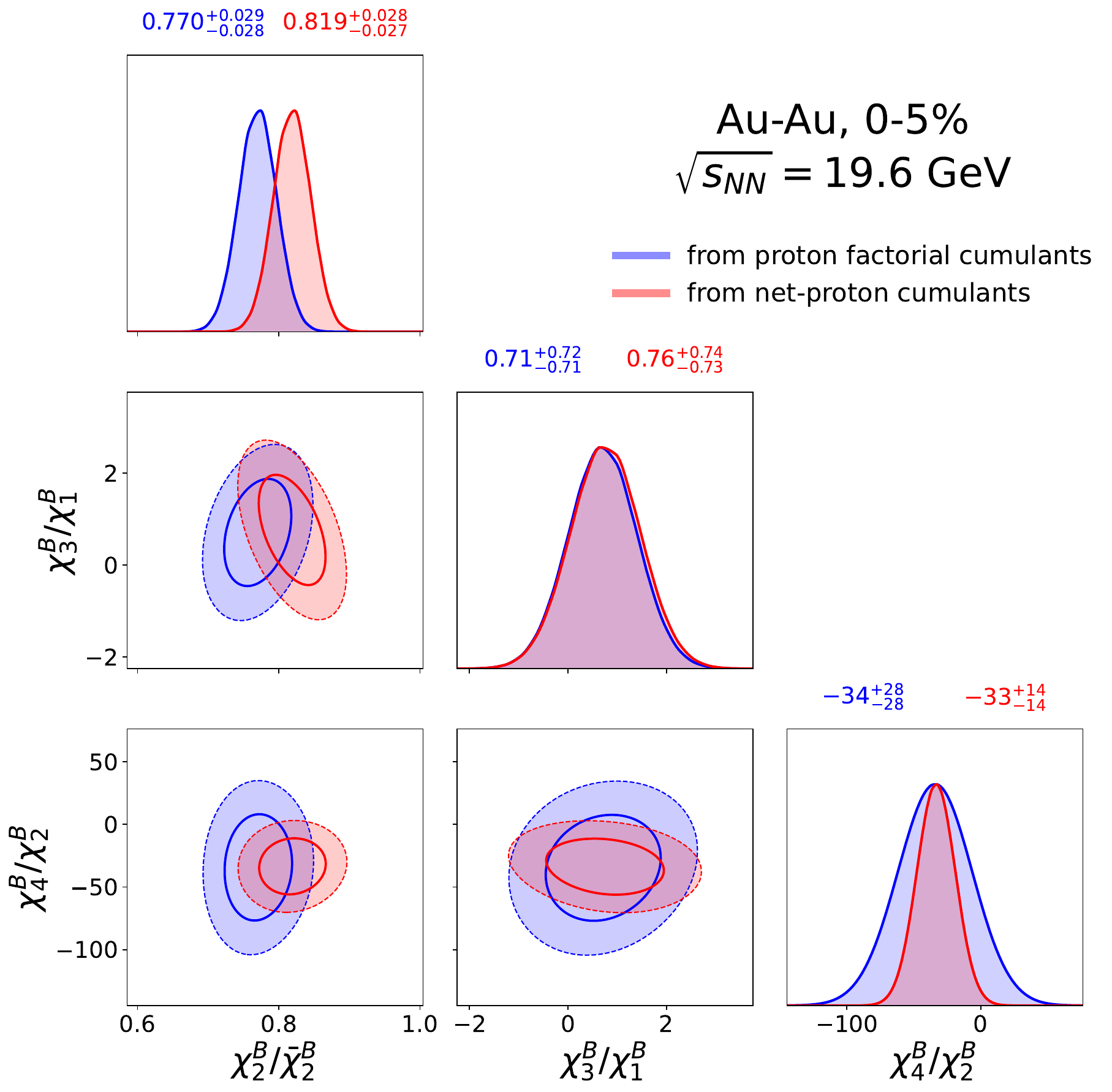}
    \caption{
    Marginal posterior distributions of the susceptibility ratios at $\sNN = 19.6$~GeV from the proton factorial cumulant fit~(blue) and the net-proton cumulant fit~(red), with the quoted values above the diagonal panels giving the medians and 68\% credible intervals. Diagonal panels: 1D marginals; off-diagonal panels: 68\%~(solid) and 95\%~(dashed) credible contours. 
    }
    \label{fig:corner}
\end{figure}

Figure~\ref{fig:corner} shows representative posterior distributions at $\sNN = 19.6$~GeV, comparing the proton factorial cumulant and net-proton cumulant fits.
The second-order ratio is tightly and symmetrically constrained and only mildly correlated with the higher-order parameters.
The fourth-order ratio exhibits the largest uncertainties, reflecting the signal dilution of the corresponding factorial cumulant; the two posteriors are mutually consistent, with the net-proton fit yielding the tighter fourth-order constraint.

\begin{figure*}[t]
    \centering
    \includegraphics[width=0.99\textwidth]{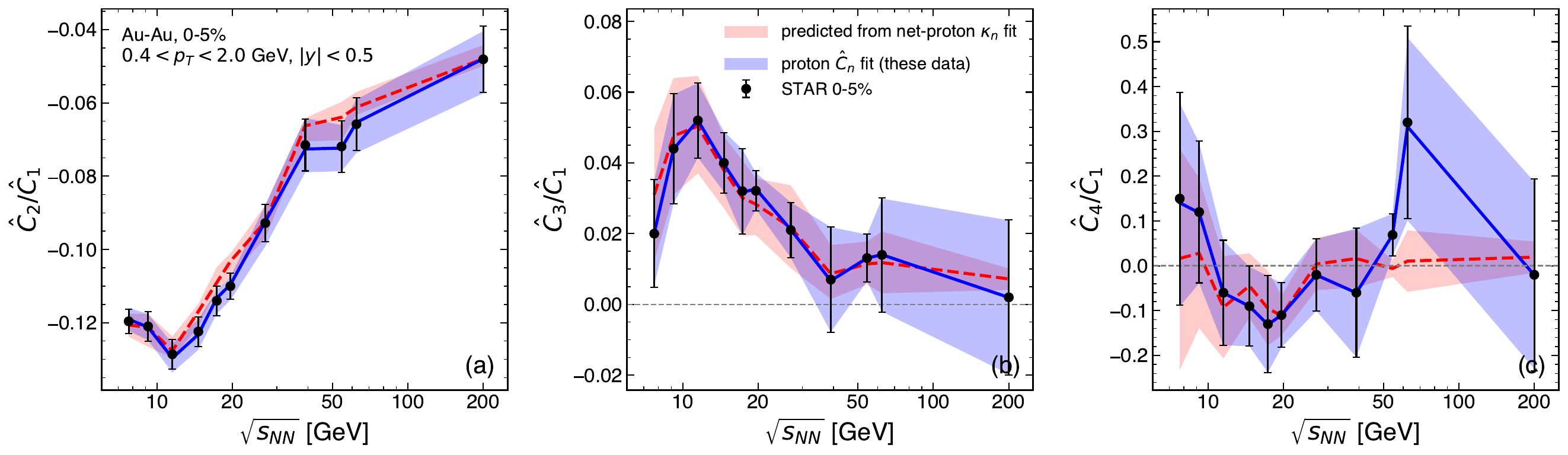}\\
    \includegraphics[width=0.99\textwidth]{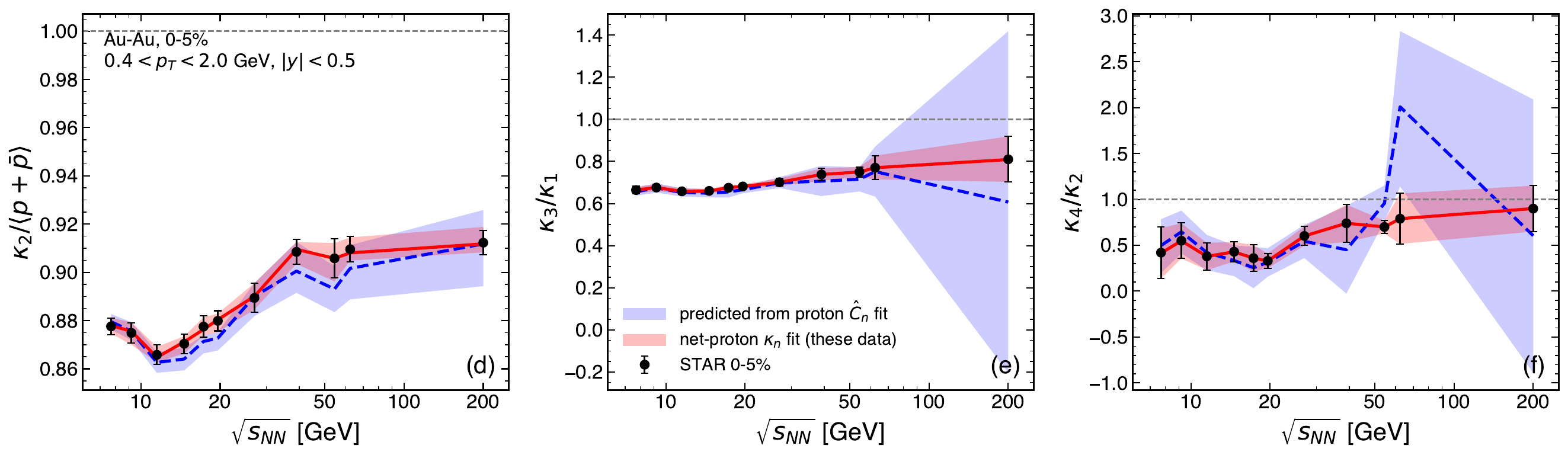}
    \caption{
    Cross-validation of the two extractions. Panels~(a)--(c): proton factorial cumulant ratios with the band from the fit to data~(blue) and the prediction from the net-proton fit~(red dashed). Panels~(d)--(f): same for the net-proton cumulant ratios, with the prediction from the proton factorial cumulant fit. STAR data from Refs.~\cite{STAR:2025zdq,STAR:2020tga,STAR:2021iop}. 
    }
    \label{fig:cross}
\end{figure*}

Figure~\ref{fig:cross} presents the cross-validation of the two extractions discussed in the main text: each observable set is compared to the band from its own fit and to the prediction obtained from the posterior of the other data set.

\section{Extracted parameter values}

\begin{table*}[!t]
\caption{
Extracted baryon number susceptibility ratios~(marginal posterior median and 68\% credible interval) in 0--5\% central Au-Au collisions, from the proton factorial cumulant and net-proton cumulant fits.
The collision energies $\sNN = 7.7$--$27$~GeV correspond to BES-II data~\cite{STAR:2025zdq}, $39$--$200$~GeV to BES-I~\cite{STAR:2020tga,STAR:2021iop}, and the 14.6~GeV data are compared to the hydrodynamic surface simulated at $\sNN = 14.5$~GeV.
\label{tab:results}
}
\begin{ruledtabular}
\begin{tabular}{lcccccc}
 & \multicolumn{3}{c}{Proton factorial cumulants} & \multicolumn{3}{c}{Net-proton cumulants} \\
$\sNN$~[GeV] & $\cc{2}/\ccb{2}$ & $\cc{3}/\cc{1}$ & $\cc{4}/\cc{2}$ & $\cc{2}/\ccb{2}$ & $\cc{3}/\cc{1}$ & $\cc{4}/\cc{2}$ \\
\hline
7.7 & $1.074 \pm 0.035$ & $-0.5 \pm 3.7$ & $79^{+107}_{-110}$ & $1.060 \pm 0.035$ & $1.9 \pm 4.2$ & $46 \pm 111$ \\
9.2 & $0.936 \pm 0.038$ & $2.7 \pm 2.4$ & $86^{+76}_{-73}$ & $0.928 \pm 0.037$ & $3.2 \pm 2.2$ & $58^{+64}_{-63}$ \\
11.5 & $0.765 \pm 0.035$ & $1.9 \pm 1.3$ & $3 \pm 53$ & $0.781 \pm 0.032$ & $1.9 \pm 1.3$ & $-9 \pm 42$ \\
14.6 & $0.731 \pm 0.033$ & $0.7 \pm 1.0$ & $-24 \pm 37$ & $0.768 \pm 0.029$ & $0.8 \pm 0.9$ & $-6 \pm 25$ \\
17.3 & $0.767 \pm 0.032$ & $0.5 \pm 1.3$ & $-40 \pm 41$ & $0.813 \pm 0.031$ & $0.8 \pm 1.0$ & $-25 \pm 27$ \\
19.6 & $0.770^{+0.029}_{-0.028}$ & $0.7 \pm 0.7$ & $-34 \pm 28$ & $0.819^{+0.028}_{-0.027}$ & $0.8 \pm 0.7$ & $-33 \pm 14$ \\
27 & $0.834 \pm 0.037$ & $0.5 \pm 0.8$ & $-3 \pm 25$ & $0.833 \pm 0.037$ & $0.6 \pm 1.0$ & $5 \pm 16$ \\
39 & $0.844 \pm 0.054$ & $-0.4 \pm 2.4$ & $-20 \pm 54$ & $0.881 \pm 0.029$ & $0.3 \pm 1.1$ & $7 \pm 23$ \\
54.4 & $0.790 \pm 0.055$ & $0.5 \pm 1.5$ & $31 \pm 20$ & $0.853 \pm 0.042$ & $0.9 \pm 0.9$ & $-0.2 \pm 7.3$ \\
62.4 & $0.816 \pm 0.056$ & $1.2 \pm 3.6$ & $133^{+83}_{-86}$ & $0.862 \pm 0.028$ & $1.2 \pm 1.8$ & $7 \pm 26$ \\
200 & $0.848 \pm 0.077$ & $-4 \pm 27$\footnote{The standard prior $\cc{3}/\cc{1} \in [-15,15]$ is too narrow, therefore, the quoted value uses a widened prior $[-60,60]$ for which the posterior was verified to saturate.
} & $-8 \pm 114$ & $0.857 \pm 0.025$ & $2.3 \pm 3.6$ & $12 \pm 19$ \\
\end{tabular}
\end{ruledtabular}
\end{table*}

\begin{table*}[!t]
\caption{
Second-order baryon number susceptibility along the chemical freeze-out line~\cite{Vovchenko:2015idt}, with $T$ and $\mu_B$ given by the freeze-out parametrization at each collision energy.
The extracted values from the two fits are normalized as $\cc{2} = (\cc{2}/\ccb{2}) \times \ccb{2}(T, \mu_B)$ with $\ccb{2}$ evaluated in the ideal QMHRG2020 model~\cite{Bollweg:2021vqf}, and are compared to the lattice QCD-based 4D-TExS equation of state~\cite{Abuali:2025tbd,MUSES_4DTExS_v1_0_4}~(cf.~Fig.~\ref{fig:chi2B} of the main text) and to the QMHRG2020 baseline itself.
\label{tab:chi2B}
}
\begin{ruledtabular}
\begin{tabular}{lcccccc}
 & & & \multicolumn{2}{c}{Extracted $\cc{2}$} & \multicolumn{2}{c}{EoS $\cc{2}$} \\
$\sNN$~[GeV] & $T$~[MeV] & $\mu_B$~[MeV] & fact.~cumulants & net-$p$~cumulants & 4D-TExS & QMHRG2020 \\
\hline
7.7 & 140 & 406 & $0.395 \pm 0.013$ & $0.390 \pm 0.013$ & $0.250 \pm 0.025$ & $0.368$ \\
9.2 & 145 & 355 & $0.294 \pm 0.012$ & $0.291 \pm 0.011$ & $0.215 \pm 0.021$ & $0.314$ \\
11.5 & 149 & 299 & $0.200 \pm 0.009$ & $0.204 \pm 0.008$ & $0.187 \pm 0.016$ & $0.261$ \\
14.6 & 151 & 246 & $0.160 \pm 0.007$ & $0.168 \pm 0.006$ & $0.165 \pm 0.011$ & $0.219$ \\
17.3 & 153 & 213 & $0.151 \pm 0.006$ & $0.160 \pm 0.006$ & $0.153 \pm 0.009$ & $0.197$ \\
19.6 & 154 & 191 & $0.142 \pm 0.005$ & $0.151 \pm 0.005$ & $0.146 \pm 0.007$ & $0.184$ \\
27 & 155 & 144 & $0.134 \pm 0.006$ & $0.134 \pm 0.006$ & $0.134 \pm 0.005$ & $0.161$ \\
39 & 156 & 103 & $0.123 \pm 0.008$ & $0.129 \pm 0.004$ & $0.126 \pm 0.003$ & $0.146$ \\
54.4 & 157 & 75 & $0.110 \pm 0.008$ & $0.119 \pm 0.006$ & $0.122 \pm 0.002$ & $0.139$ \\
62.4 & 157 & 66 & $0.112 \pm 0.008$ & $0.118 \pm 0.004$ & $0.121 \pm 0.002$ & $0.137$ \\
200 & 157 & 21 & $0.111 \pm 0.010$ & $0.112 \pm 0.003$ & $0.118 \pm 0.001$ & $0.131$ \\
\end{tabular}
\end{ruledtabular}
\end{table*}

Table~\ref{tab:results} lists the extracted susceptibility ratios at all collision energies from both fits.
Table~\ref{tab:chi2B} compares the extracted second-order susceptibility $\cc{2}$, evaluated along the chemical freeze-out line~\cite{Vovchenko:2015idt}, with the 4D-TExS and QMHRG2020 values.

\section{Two-baryon correlation dominance and stage-by-stage decomposition}
\label{sec:additional}

\begin{figure*}[t]
    \centering
    \includegraphics[width=0.95\textwidth]{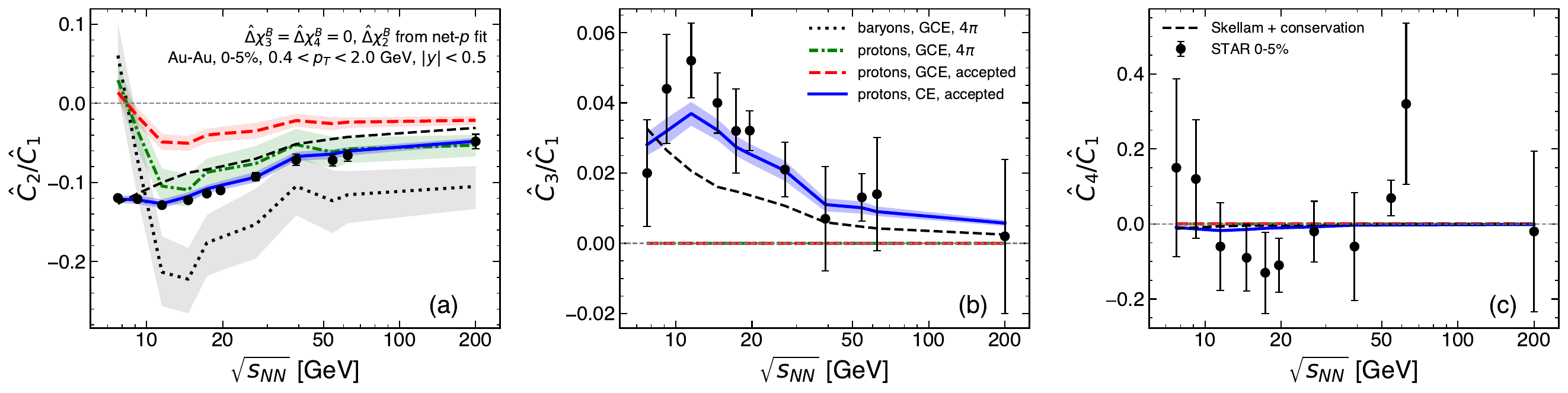}
    \caption{
    Stage-by-stage decomposition of the proton factorial cumulant ratios in the two-particle-correlation scenario, including the fourth order.
    The grand-canonical stages of $\cumc{3}/\cumc{1}$ and $\cumc{4}/\cumc{1}$ vanish identically in this scenario; the measured signals emerge entirely from the exact-conservation step.
    The black dashed line shows conservation acting on the uncorrelated Skellam baseline, ${\bs \gamma} = (1,1,1)$: without the energy dependence of $\cc{2}/\ccb{2}$ it yields a monotonic $\cumc{3}/\cumc{1}$ and nearly vanishing $\cumc{4}/\cumc{1}$.
    \label{fig:fcstages}
    }
\end{figure*}
Figure~\ref{fig:fcstages} provides the stage-by-stage decompositions of the proton factorial cumulant ratios in the scenario retaining only the irreducible two-baryon contribution, complementing Fig.~\ref{fig:stages} of the main text.
Each successive stage -- restriction to protons, kinematic acceptance, and exact baryon conservation -- modifies $\cumc{2}/\cumc{1}$ substantially, and only the full chain describes the data.
Higher-order factorial cumulants vanish identically in the grand-canonical calculation here because irreducible three- and four-baryon correlations are neglected.
The non-zero values arise from the baryon conservation correction, with a magnitude driven by the overall accepted fraction of baryons and influenced by the two-baryon correlations.

\end{appendix}

\end{document}